\documentclass[aps,prd,twocolumn,floatfix,altaffilletter,superscriptaddress,preprintnumbers, tightenlines,showpacs,showkeys,nofootinbib,notoccite]{revtex4-1}
\usepackage[utf8]{inputenc}
\usepackage[colorlinks=true,citecolor=blue,linkcolor=blue]{hyperref}
\usepackage[normalem]{ulem}
\usepackage{amsmath,amssymb, mathrsfs,esvect}
\usepackage{epsfig}
\usepackage{graphicx}               
\usepackage{url}
\usepackage{color}
\usepackage{slashed}
\usepackage{multirow}
\usepackage{placeins}
\usepackage[dvipsnames]{xcolor}
\usepackage{epstopdf}
\usepackage{soul}
\usepackage{tikz}
\usepackage[capitalise, english]{cleveref}
\usepackage{siunitx}
\usepackage{xspace}
\usepackage{booktabs}
\usetikzlibrary{trees}
\usetikzlibrary{decorations.pathmorphing}
\usetikzlibrary{decorations.markings}

\newcommand\myshade{80}
\colorlet{mylinkcolor}{ForestGreen}
\colorlet{mycitecolor}{Red}
\colorlet{myurlcolor}{violet}

\hypersetup{
	linkcolor  = mylinkcolor!\myshade!black,
	citecolor  = mycitecolor!\myshade!black,
	urlcolor   = myurlcolor!\myshade!black,
	colorlinks = true
}

\definecolor{jblue}{RGB}{20,50,100}
\definecolor{wred}{RGB}{217,0,56}
\definecolor{white}{RGB}{255,255,255}
\definecolor{forestgreen}{HTML}{228B22}

\allowdisplaybreaks
\definecolor{lime}{HTML}{A6CE39}
\DeclareRobustCommand{\orcidicon}{\hspace{-1mm}
	\begin{tikzpicture}
		\draw[lime, fill=lime] (0,0) 
		circle [radius=0.16] 
		node[white] {{\fontfamily{qag}\selectfont \tiny \,ID}};
		\draw[white, fill=white] (-0.0525,0.095) 
		circle [radius=0.007];
	\end{tikzpicture}
	\hspace{-3mm}
}

\foreach \x in {A, ..., Z}{\expandafter\xdef\csname orcid\x\endcsname{\noexpand\href{https://orcid.org/\csname orcidauthor\x\endcsname}
		{\noexpand\orcidicon}}
}

\newcommand{\Msun}{{\rm M}_\odot}

\begin{document}


\title{Extreme Ultraviolet Microlensing of Primordial Black Hole Dark Matter \\ in the Asteroid-Mass Window}

\author{Soorna Choheili\orcidB}
\affiliation{Department of Physics, Engineering Physics, and Astronomy, Queen's University, Kingston, ON, K7L 3N6, Canada}
\affiliation{Arthur B. McDonald Canadian Astroparticle Physics Research Institute, Kingston, ON, K7L 3N6, Canada}

\author{Anupam Ray\orcidC}
\affiliation{Department of Physics, Engineering Physics, and Astronomy, Queen's University, Kingston, ON, K7L 3N6, Canada}
\affiliation{Arthur B. McDonald Canadian Astroparticle Physics Research Institute, Kingston, ON, K7L 3N6, Canada}
\affiliation{Perimeter Institute for Theoretical Physics, Waterloo, Ontario, Canada}

\author{Joseph Bramante\orcidA}
\affiliation{Department of Physics, Engineering Physics, and Astronomy, Queen's University, Kingston, ON, K7L 3N6, Canada}
\affiliation{Arthur B. McDonald Canadian Astroparticle Physics Research Institute, Kingston, ON, K7L 3N6, Canada}

\date{\today}

\begin{abstract}
Primordial black holes (PBHs) in the asteroid-mass range $[10^{-16}$--$10^{-11}]\,M_\odot$ could comprise the majority of dark matter, and this possibility remains stubbornly unexplored. In this work, we propose the use of extreme ultraviolet (EUV) microlensing as a novel probe of asteroid-mass PBH dark matter. This work implements the full microlensing formalism including wave optics, finite source effects, and energy-averaged magnification,  and systematically identify optimal targets across all EUVE source classes. Using archival data from the Extreme Ultraviolet Explorer (EUVE),  we derive a limit $f_{\rm PBH} \gtrsim 100$ over the  mass range $[5 \times 10^{-13}$--$10^{-11}]\,M_\odot$. While this bound on the PBH fraction of dark matter exceeds unity and is thus not constraining, it establishes that EUV microlensing is a viable observational channel. Looking ahead, we show that the proposed CASTOR space telescope could improve on present EUVE sensitivity by almost three orders of magnitude, reaching $f_{\rm PBH} \approx 0.35$ with a one-year exposure, which would be sufficient to test asteroid-mass PBHs as the sole dark matter component. Hence, there is a compelling case for dedicated microlensing time on future UV/EUV space telescopes to potentially discover PBH dark matter.
\end{abstract}

\maketitle

\section{Introduction}
\label{sec:intro}
The microscopic identity of dark matter (DM), which accounts for approximately 27\% of the total energy density of the Universe, remains one of the foremost open questions in fundamental physics~\cite{Bozorgnia:2024pwk,Bramante:2026wzh,Carney:2022gse}. Among the various DM candidates, primordial black holes (PBHs), possibly formed in the early Universe through the gravitational collapse of large density fluctuations generated during inflation, are particularly 
compelling~\cite{1966AZh....43..758Z,Hawking:1971ei,Carr:1974nx,0912.5297,CarrKohri2021,CarrKuhnelSandstad2016,Carr:2020xqk,Green2021,Escriva:2022duf, Carr:2026hot,Ray:2022rfz}.

Extensive observational campaigns using a diverse range of astrophysical and cosmological probes have placed significant constraints on the PBH abundance across a wide mass range (see recent reviews~\cite{0912.5297,CarrKuhnelSandstad2016,
Carr:2020xqk,CarrKohri2021,Green2021,Escriva:2022duf,Carr:2026hot}). At the lightest end of the mass spectrum, $M_{\rm PBH} \in [10^{-18}, 10^{-16}]\,M_{\odot}$, commonly referred to as the ultralight-mass window, PBHs are subject to stringent 
constraints from Hawking radiation. Such light PBHs can emit several Standard Model particles through Hawking evaporation, and the non-detection of these particles has constrained the dark 
matter fraction of light PBHs~\cite{0912.5297, 1807.03075, 1906.04750, 1906.07740, 
1906.09994, 1912.01014, Ballesteros:2019exr, 2004.00627, 2007.01589, Ray:2021mxo, 
2010.04797, 2007.07739, 2009.11837, 2007.05677, Ray:2021mxo, 2210.00072, 
2203.14979, 2404.17119, 2406.11949, Koechler:2026, Li:2026}. In addition, particles emitted through Hawking evaporation will alter the thermal and ionization history of the Universe, leaving observable imprints on cosmological measurements, which further constrains the light PBH abundance~\cite{1612.07738, 2107.02190, 
1801.01871, 1907.06485, 2002.00898, 2011.12244, 1803.09390, 2107.12358, 2108.13256, 
2112.10794, Khan:2025StrongerConstraints}. Here it is worth-noting that PBHs with masses below approximately $10^{-18}\,M_{\odot}$ have lifetimes shorter than the present age of the Universe, precluding them from constituting the present-day dark matter~\cite{Page:1976df,Page:1976ki,
MacGibbon:2007yq}.

Moving to higher masses, the \textit{asteroid}-mass window, $M_{\rm PBH} \in 
[10^{-16},10^{-11}]\,M_{\odot}$, remains essentially unexplored by existing 
PBH searches, leaving open the possibility that PBHs in this mass range could constitute the entirety of dark matter. (More massive PBHs are further constrained by optical lensing, dynamical friction, GW observations, accretion signatures etc, see \textit{e.g.}~\cite{Green2021,Escriva:2022duf} for recent reviews.) A new search method for asteroid-mass PBHs is the central focus of our work. Several astrophysical constraints have been proposed in the asteroid-mass range~\cite{Barnacka:2012bm,Capela:2013yf,
Graham:2015apa}, however, these have been subsequently weakened by follow-on work in ~\cite{Katz2018,MonteroCamacho:2019jte,Smyth:2019whb,Genolini:2020}. 
Other approaches have also been explored, including searches for Hawking-radiated MeV photon signatures~\cite{2010.04797,Ray:2021mxo} and gamma-ray burst lensing parallax~\cite{Jung:2020jrs,Fedderke:2024picolensing}, which are plausible future avenues for finding asteroid-mass PBHs.

In this work, we propose gravitational microlensing in the extreme ultraviolet (EUV) band as a novel probe of asteroid-mass PBH dark matter. Building on recent work~\cite{Bai2019,Tamta2024} that explored microlensing of X-ray pulsars as a probe of asteroid-mass PBH dark matter, we investigate the potential of EUV observations to provide a complementary search for asteroid-mass PBHs. To establish the viability of EUV searches, we analyze archival data from the Extreme Ultraviolet Explorer (EUVE)~\cite{BowyerMalina1991}, which operated from 1992 to 2001 in the 70--760~\AA\ (16--177~eV) band, and derive the \textit{first} microlensing constraints from EUV observations. A key point is that the EUV band accesses photon energies intermediate between optical and X-ray, enabling sensitivity to PBH masses in the range $[5\times10^{-13}$--$10^{-11}]\,M_\odot$, which is a window that has not been explored by any microlensing technique. We assess the prospects for improved PBH sensitivity with the proposed CASTOR telescope~\cite{Cote2019,CASTOR_SPIE2025}, which we show could discover asteroid-mass PBHs across a decade in mass with a one-year exposure of a known pulsar. Altogether, our results establish UV microlensing as a puissant new probe of asteroid-mass PBH dark matter.

PBHs are not the only kind of dark matter that UV microlensing could discover. There is a broader category of dark compact objects (DCOs) that are also dark matter candidates, and may populate the halo with asteroid-mass microlensing targets. Models predicting these include axion miniclusters~\cite{Kolb:1993zz,Fairbairn2017}, boson stars~\cite{Liebling:2012fv,Acevedo:2022gug}, Q-ball dark matter~\cite{Kusenko:1997si}, composite dark matter comprised of fermions~\cite{Bai:2018dxf,BaiLongLu2020,Bramante:2026wzh,Acevedo:2024lyr,Bleau:2025klr}, and objects formed in a dissipative dark matter sector~\cite{Fan:2013yva,Shandera:2018xkn,Bramante:2023ddr,Bramante:2024pyc}. Unlike PBHs, the radii of these DCOs is not fixed by their masses, and so their microlensing properties will depend on both mass and compactness. DCOs may also be sought via gravitational waves~\cite{Giudice:2016zpa} and a stellar dimming effect~\cite{Bramante:2024hbr}. We will find there is UV microlensing sensitivity for finite-size DCOs in Sec.~\ref{sec:massradius}.

The rest of this paper explores methods and prospects for UV microlensing of PBHs and DCOs. In Sec.~\ref{sec:theory} we detail the microlensing formalism and specifically emphasize wave optics and finite source effects at EUV energies. There are important EUVE source classes and optimal targets laid out in Sec.~\ref{sec:sources}. Sec.~\ref{sec:detection} describes the analytical framework for discriminating microlensing events from backgrounds in UV data. We detail DM lensing event rates and our limit-setting procedure in Sec.~\ref{sec:rates}. Sec.~\ref{sec:results} presents new results, including limits set using EUVE data and a comparison with the previous X-ray microlensing PBH studies. Finally, in Sec.~\ref{sec:discussion} we present our conclusions.

\section{PBH Microlensing formalism}
\label{sec:theory}
We briefly review the theoretical framework underlying gravitational microlensing 
by a point-mass lens, highlighting the three ingredients that distinguish the EUV 
regime from standard treatments: wave optics corrections, finite source size 
suppression, and energy-averaged magnification.

\paragraph{Lens geometry.}
We consider a the PBH to be a point mass $M$ at distance $D_L$ from the 
observer, in front of a background source at distance
$D_S$~\cite{NarayanBartelmann1996}. Following the literature, we define the fractional lens distance 
$x \equiv D_L/D_S$, with corresponding Einstein radius
\begin{equation}
r_E = \sqrt{\frac{4GM\,x(1-x)\,D_S}{c^2}} \,,
\label{eq:rE}
\end{equation}
which can be manipulated to
\begin{align}
r_E = 115~{\rm km}
&\left(\frac{M}{10^{-12}M_\odot}\right)^{1/2}
\left(\frac{D_S}{290~{\rm pc}}\right)^{1/2} \nonumber \\
&\times\left(\frac{x(1-x)}{0.25}\right)^{1/2},
\end{align}
with the normalization chosen for our primary EUV target, PSR~B0656+14 
($D_S = 290$~pc). The rest of this analysis will be done with respect to $r_E$. For convenience, we will define $y$ to be the impact parameter, and $R_S$ as 
the physical radius of the emission region. The value $a_S(x) \equiv x R_S / r_E$ then evaluates to be the source radius projected onto the lens plane. The parameter $a_S$ controls how extended the sources appear relative to the Einstein ring and, as we discuss 
below, is the dominant factor determining whether a given source class is a 
viable microlensing target.

\paragraph{Wave optics and the minimum accessible mass.}
Standard microlensing treatments adopt the geometric optics approximation, valid 
when the photon wavelength is much smaller than the Schwarzschild radius of the 
lens. For asteroid-mass PBHs observed in the EUV band this condition is not 
generically satisfied, requiring a full wave-optics treatment. The transition 
between regimes is captured by the dimensionless parameter
\begin{equation}
w \equiv \frac{4GM E_\gamma}{\hbar c^3} = \frac{2R_{\rm Schw}\,E_\gamma}{\hbar c} \,,
\label{eq:w}
\end{equation}
which measures the ratio of the photon energy to the inverse light-crossing time of the Schwarzschild radius. When $w \gg 1$ we recover the geometric optics limit, whereas for $w \lesssim 1$ diffraction suppresses the magnification. Setting $w \sim 1$ therefore gives the smallest PBH mass that can be detected at a given photon energy, through
\begin{equation}
E_\gamma = 33.5~{\rm keV}\left(\frac{10^{-15}M_\odot}{M}\right)w \,,
\label{eq:Egamma}
\end{equation}
which gives $M_{\rm min} \approx 3 \times 10^{-13}\,M_\odot$ for EUVE ($E_{\rm max} \approx 124$~eV). Instruments that detect higher-energy photons can reach lower masses: $\sim 3 \times 10^{-15}\,M_\odot$ for NICER ($E_{\rm max} = 12$~keV) and $\sim 1 \times 10^{-15}\,M_\odot$ for STROBE-X ($E_{\rm max} = 30$~keV). X-ray instruments will thus be sensitive to lighter PBHs, 
while EUV microlensing can detect the heavier end of the 
asteroid-mass PBH window.

\paragraph{Magnification.}
With $w$ and $a_S$ defined, the magnification accounting for both wave optics 
and a finite Gaussian source 
profile is~\cite{NakamuraDeguchi1999,MatsunagaYamamoto2006,TakahashiNakamura2003}
\begin{align}
\mu(w,y,a_S) &= \frac{e^{-y^2/(2a_S^2)}}{a_S^2}\,\frac{\pi w}{1-e^{-\pi w}} 
\nonumber \\
&\quad \times \int_0^\infty dz\,z\,e^{-z^2/(2a_S^2)}\,I_0\!\left(\frac{yz}{a_S^2}
\right) \nonumber \\
&\quad \times \left|{}_1F_1\!\left(\frac{iw}{2},1;\frac{iwz^2}{2}\right)\right|^2 \,,
\label{eq:mu_full}
\end{align}
where ${}_1F_1$ is the confluent hypergeometric function of the first kind and 
$I_0$ is the zeroth-order modified Bessel function. In the geometric optics limit
($w \gg 1$), the photon wavelength is much smaller than the Schwarzschild radius of the lens, so wave effects can be neglected and the magnification no longer depends on photon energy. Eq.~(\ref{eq:mu_full}) then reduces to the familiar point-source result
\begin{equation}
\mu_{\rm geo}(y) = \frac{y^2 + 2}{y\sqrt{y^2 + 4}} \,,
\label{eq:mu_geo}
\end{equation}
which we use as an analytic shortcut at large $w$, where the ${}_1F_1$ series converges slowly. The magnification also depends on the source size through $a_S$. When $a_S \ll 1$, the source is effectively point-like and $\mu$ approaches its full single-lens value. When $a_S \gtrsim 1$, the magnification varies across the source disk: regions close to the lens are strongly magnified, while more distant regions are barely magnified. When the magnification is averaged over the whole disk, the result therefore ends up close to one~\cite{WittMao1994}. In other words, a large source barely brightens, even when the lens passes right across its center. In other words, a huge source rarely illuminates even when the lens passes directly through its center. This effect is referred to as finite-source suppression. It occurs because just the region of the disk nearest the lens is substantially magnified, and that portion contains only a small fraction of the total light. The larger the source relative to the Einstein radius, the smaller this proportion grows; hence, a source many Einstein radii across can only be brightened by a few percent. The effect is completely geometric and has nothing to do with wave optics. Sec.~\ref{sec:sources} determines whether EUV sources are suitable for microlensing. 

Because $w$ is proportional to photon energy, magnification varies with energy, yet the detector captures photons across a broad band rather than at a single energy. The EUVE band ranges from 10 to 124 eV, resulting in a significant variation in $w$. A single lens can thus exhibit wave effects for the softest photons while operating virtually geometrically on the hardest. The detector cannot distinguish between these photons, therefore it reports the magnification averaged across the band, with each energy weighted by the number of photons observed. For a source with spectrum $F(E)$ viewed by a telescope with effective area $A(E)$, this band-averaged magnification is \begin{equation} \bar{\mu}_E(y, a_S) = \frac{\displaystyle\int_{E_{\rm min}}^{E_{\rm max}} dE\, A(E)\,F(E)\, \mu(w(E), y, a_S)}{\displaystyle\int_{E_{\rm min}}^{E_{\rm max}} dE\, A(E)\,F(E)} 
\,,
\label{eq:muE}
\end{equation}
with $w(E)$ defined by Eq.~(\ref{eq:w}). The weight $A(E)F(E)$ represents the amount of photons observed at each energy, where $F(E)$ determines how much light the source emits and $A(E)$ determines how much light the telescope catches. The average is thus dominated by energies at which the source is bright and the telescope is sensitive. The denominator guaranties that $\bar{\mu}_E = 1$ when there is no lensing. In the geometric optics limit, $\mu$ is the same at all energies and $\bar{\mu}_E$ decreases to $\mu$, therefore the weighting is only relevant when wave effects arise within the band. We utilize $\bar{\mu}_E$ throughout Sec.~\ref{sec:rates}, and we apply the detection threshold of Sec.~\ref{sec:detection} to it, because the band-averaged brightening must rise above the noise. 

\section{UV Microlensing Source selection}
\label{sec:sources}

The primary challenge for EUV microlensing is a lack of photons. An event is only detectable if it lifts the source clearly above the random fluctuations in its photon counts, and these fluctuations are larger, relative to the mean, for fainter sources. Counting noise alone gives fluctuations of about $1/\sqrt{N}$ for $N$ photons per bin, or 10--20\% for 25--100 photons, and variability from the source and background adds to this. The fainter the source, the stronger the magnification an event needs in order to stand out.

There are two reasons EUV sources are faint. First, few objects emit most of their light in the EUV. Ordinary stars are brightest in the optical, while most compact X-ray sources peak at much higher energies. The most promising targets are hot, compact objects such as young neutron stars and white dwarfs, whose temperatures of tens of eV place much of their thermal emission near the EUV band.

Most of this absorption is due to neutral hydrogen. Photons near 100~eV easily ionize hydrogen atoms, with a cross section of $\sigma \sim 6 \times 10^{-18}$~cm$^2$, far larger than at X-ray energies. How much light gets through depends on the hydrogen column density $N_H$, the number of hydrogen atoms per unit area along the line of sight. A column of only $N_H \sim 10^{18}$~cm$^{-2}$ gives an optical depth $\tau_{\rm ISM} \sim 1$, which cuts the flux by a factor of $e$. Since the column grows with distance and the flux falls exponentially with $\tau_{\rm ISM}$, absorption becomes severe quickly. Doubling the distance along a uniform sightline, for example, costs another factor of $e$, on top of the usual inverse-square falloff.

That EUV sources are visible at interstellar distances at all is due to the Sun's location inside the Local Bubble, a region of unusually thin gas. Even here, a column of $N_H \sim 10^{18}$~cm$^{-2}$ accumulates within about 100--300~pc, depending on direction, and beyond the bubble the gas is denser still.

The result is an effective horizon for EUV astronomy, which limits microlensing in two ways. First, it rules out sources farther than a few hundred parsecs. Second, it limits the event rate even for the sources that remain, because a nearby source has little dark matter in front of it. Since the lensing optical depth grows roughly as $D_S^2$ for nearby sources, a source at 300~pc offers about a hundred times fewer chances to be lensed than one at 3~kpc would. X-rays face no such limit. The photoionization cross section falls steeply with photon energy, so keV X-rays cross many kiloparsecs of the Galaxy with little absorption. This is why the X-ray analysis of Ref.~\cite{Tamta2024} can use sources as distant as SMC~X-1, at 64~kpc, toward which the lensing optical depth is roughly $10^5$ times larger than toward a source a few hundred parsecs away.

Within this horizon, the most important property of a source is its size. As Sec.~\ref{sec:theory} showed, the magnification is strongly reduced once the source size parameter $a_S$, the source radius projected onto the lens plane and measured in Einstein radii, is larger than about one. Because $a_S$ is measured in Einstein radii, it depends on the lens mass and on where the lens sits along the line of sight, as well as on the size of the source. A useful source must therefore be small compared with $r_E$ at the lens masses of interest. Since $r_E \propto \sqrt{M}$, a source that is too large for light PBHs may still be small enough for heavier ones. As an example, consider a PBH of mass $M = 10^{-11}\,M_\odot$ halfway to a source at $D_S = 290$~pc ($x = 0.5$). Its Einstein radius is $r_E \approx 364$~km, which gives:
\begin{itemize}
\item Neutron star ($R_S \approx 13$~km): $a_S \approx 0.02$
\item White dwarf ($R_S \approx 7000$~km): $a_S \approx 10$
\item sdO/sdB star ($R_S \approx 10^5$~km): $a_S \approx 140$
\item Main-sequence star ($R_S \gtrsim 10^6$~km): $a_S \gtrsim 1400$
\end{itemize}
Only neutron stars have $a_S \ll 1$, so at this mass they are the only sources that can be magnified without suppression. For a white dwarf, $a_S$ falls to about one only at $M \sim 10^{-9}\,M_\odot$, a hundred times heavier. As Sec.~\ref{sec:results} shows, these masses lie well inside the region already excluded by Subaru-HSC, so white dwarfs do not extend the reach of an EUV search. Hence, the primary target of our analysis is thermally-emitting neutron stars, since these provide the best prospects for UV-lensing of PBHs. Neutron stars are also well-established laboratories for dark matter physics more broadly~\cite{Bramante:2023djs,Baryakhtar:2022hbu}.

This can be confirmed quantitatively, by computing the full energy-averaged 
magnification $\bar{\mu}_E$ from Eq.~(\ref{eq:muE}), using EUVE's effective 
area, at a representative PBH mass $M = 10^{-10}\,M_\odot$, impact parameter $y = 0.1$, 
and $x = 0.5$. For PSR~B0656+14 we find $\bar{\mu}_E \approx 10$, which is well above a 
reasonable magnification detection threshold. For the hot white dwarf HZ~43, $\bar{\mu}_E 
\approx 1.02$, indicating a very small magnification. For the 
sdO/sdB and main-sequence stars, $\bar{\mu}_E < 1.01$. Hence we see that with presently available UV telescope data, only neutron stars yield large enough magnifications for microlensing detection. 

We have nevertheless for completeness compiled a list of EUV-bright neutron stars observed by the EUVE  Deep Survey (DS) spectrometer~\cite{Bowyer1994,Bowyer1996}, in addition to hot  white dwarfs and early-type (O/B) stars. Their 
microlensing-relevant parameters are summarized in Table~\ref{tab:sources}.

\begin{table*}[t]
\centering
\begin{tabular}{lcccccc}
\hline\hline
Source & $D_S$ (pc) & $R_S$ (km) & Mean E (eV) & Rate (cts/s) & $\sigma_B/B$ & $a_S$ \\
\hline
\multicolumn{7}{l}{\textit{Neutron stars}} \\[2pt]
PSR B0656+14 & 290 & 13 & 65 & 2.5 & 0.50 & 0.02 \\
Geminga & 250 & 12 & 50 & 0.30 & 0.60 & 0.02 \\
PSR B1055$-$52 & 350 & 12 & 75 & 0.15 & 0.80 & 0.02 \\
\hline
\multicolumn{7}{l}{\textit{White dwarfs}} \\[2pt]
HZ~43 & 68 & 7000 & 35 & 3.0 & 0.11 & 20 \\
GD~153 & 63 & 7200 & 34 & 1.5 & 0.04 & 21 \\
Sirius~B & 2.6 & 5800 & 25 & 4.0 & 0.02 & 84 \\
MCT~0455$-$2812 & 70 & 7500 & 32 & 2.5 & 0.03 & 21 \\
\hline
\multicolumn{7}{l}{\textit{O/B stars}} \\[2pt]
$\zeta$~Oph & 112 & $5.9 \times 10^6$ & 2.9 & 2.0 & 0.06 & $1.3 \times 10^4$ \\
$\zeta$~Pup & 335 & $1.4 \times 10^7$ & 3.5 & 0.8 & 0.08 & $1.8 \times 10^4$ \\
$\epsilon$~CMa & 132 & $9.7 \times 10^6$ & 2.5 & 5.0 & 0.04 & $2.0 \times 10^4$ \\
\hline\hline
\end{tabular}
\caption{EUV source candidates for microlensing. Listed are the source distance  $D_S$, emission region radius $R_S$, characteristic EUV photon energy in eV,  EUVE Deep Survey count rate (in the 70--760~\AA\ band, i.e., 16--177~eV),  intrinsic fractional variability $\sigma_B/B$ (for HZ~43, the value measured from its light curve at 30~s binning, which is dominated by photon counting noise), and the source size parameter $a_S \equiv x R_S / r_E$, the physical source radius projected onto the lens plane in units of the Einstein radius, evaluated at $M = 10^{-11}\,M_\odot$ and $x = 0.5$. Count rates are characteristic values from the EUVE Bright  Source List and DS spectrometer observations, updated where more recent 
calibrations exist, and should be treated as order-of-magnitude estimates 
appropriate for sensitivity scaling. Only neutron stars achieve $a_S < 1$, 
the regime in which the source is effectively point-like and the magnification 
is not strongly suppressed.}
\label{tab:sources}
\end{table*}

Among the neutron stars in Table~\ref{tab:sources}, PSR~B0656+14 provides the best combination of count rate and photometric stability. Geminga and PSR~B1055$-$52 lie at similar or slightly greater distances, which slightly increases the optical depth. However, their count rates are 8--17$\times$ lower and their light curves are much more variable, so they need much larger magnifications to register an event, and this outweighs the small gain in optical depth. Their fractional variabilities of $\sigma_B/B \approx 0.60$ and $0.80$ give threshold magnifications $\bar{\mu}_{\rm thresh} = 7.0$ and $17$, respectively, compared with $\bar{\mu}_{\rm thresh} = 5.0$ for PSR~B0656+14 
(Fig.~\ref{fig:lightcurve}). We therefore choose PSR~B0656+14 as our main target. It has a thermal EUV spectrum with $k_B T \approx 65$~eV, lies 290~pc away, and its EUV emission comes from a region about the size of a neutron star's hot polar cap ($R_S \approx 13$~km)~\cite{Posselt2023}. The EUVE DS accumulated  approximately 50~hours of net exposure on this source.

\section{UV Microlensing Events}
\label{sec:detection}

We identify candidate microlensing events as runs of $N_{\rm consec}$ consecutive  time bins in which the count rate exceeds $B + N_\sigma \sigma_B$, where $B$ is 
the mean count rate and $\sigma_B$ its standard deviation. The thresholds 
$N_{\rm consec}$ and $N_\sigma$ are chosen so that the probability of a false positive from statistical fluctuations satisfies $P_{\rm false} \ll 
t_{\rm bin}/t_{\rm exp}$, ensuring that spurious detections are negligible over the full exposure, similar to the criterion used in~\cite{Bai2019}.

To keep the number of false detections low, we require the flux to stay above the $N_\sigma$ threshold for $N_{\rm consec}$ consecutive bins. If the fluctuations are Gaussian, the probability that noise alone produces such a run is
\begin{equation}
P_{\rm false} = \left[1 - \Phi(N_\sigma)\right]^{N_{\rm consec}} \,,
\label{eq:Pfalse}
\end{equation}
where $\Phi(x) = \frac{1}{2}\left[1 + {\rm erf}(x/\sqrt{2})\right]$ is the Gaussian cumulative distribution function. For $N_\sigma = 3$ and $N_{\rm consec} = 3$, this gives $P_{\rm false} = 2.5 \times 10^{-9}$ per bin. A 60-day exposure divided into $t_{\rm bin} = 0.1$~s bins contains $t_{\rm exp}/t_{\rm bin} \approx 5 \times 10^{7}$ bins, so we expect only $\approx 0.1$ false events in total, comfortably fewer than one.

In practice, the biggest difference between X-ray and EUV microlensing lies in the photon count rate. SMC~X-1 observed with NICER gives $B \approx 204$~counts/s, or $\sim 20$ counts in each 0.1~s bin. PSR~B0656+14 observed with EUVE gives only $B \approx 2.5$~cts/s, so we use much coarser 30~s bins to collect enough counts per bin. Even then the light curve is noisy, with $\sigma_B/B \approx 0.50$, and most of this scatter comes from changes in the source and background rather than from photon counting. For the faintest neutron stars (Geminga and PSR~B1055$-$52), a bin can contain fewer than $\sim 25$ counts. The Gaussian formula of Eq.~(\ref{eq:Pfalse}) is then no longer accurate, so for these sources we use Poisson statistics instead,
\begin{equation}
P_{\rm false}^{\rm Pois} = \left[1 - F_{\rm Pois}\!\left(\lfloor \lambda + 
n_\sigma\sqrt{\lambda} \rfloor;\, \lambda\right)\right]^{N_{\rm consec}} \,,
\label{eq:Ppoisson}
\end{equation}
where $\lambda = B \cdot t_{\rm bin}$ is the expected number of counts per bin and $F_{\rm Pois}$ is the Poisson cumulative distribution function. For these sources, switching to Poisson statistics changes $P_{\rm false}$ by 10--30\%, and we carry this correction through the event rate calculation. These criteria together set the smallest magnification we are able to detect. An event is flagged once the observed flux climbs above $B + N_\sigma\sigma_B$. If we measure the magnification relative to a reference level $B + n_\sigma\sigma_B$ for the unlensed flux, the threshold magnification is
\begin{equation}
\bar{\mu}_{\rm thresh} = \frac{1 + N_\sigma \sigma_B/B}{1 + n_\sigma \sigma_B/B} \,.
\label{eq:muthresh}
\end{equation}

As in Ref.~\cite{Tamta2024}, we choose $n_\sigma = -1$, which puts the reference level one standard deviation below the mean flux. A lower reference level means that a larger magnification is needed to reach the threshold, so this choice keeps our limits conservative. The threshold rises steeply as the source gets noisier. For PSR~B0656+14, with $\sigma_B/B \approx 0.50$ and $N_\sigma = 3$, we obtain $\bar{\mu}_{\rm thresh} \approx 5.0$, while SMC~X-1 observed with NICER requires only $\bar{\mu}_{\rm thresh} = 2.5$. In other words, the noisier a source is, the more strongly it must be magnified before a lensing event can be distinguished from an ordinary fluctuation.

Microlensing also offers a simple way to rule out other brief brightenings, whether they originate in the source or in the instrument. In the geometric optics regime, a lens brightens photons of all energies by the same factor, whereas X-ray flares, trapped electrons, and detector artifacts usually appear in only one energy band. We exploit this by dividing the EUVE photons into two energy bands and comparing their count rates during any candidate event. For PSR~B0656+14, whose spectrum has $k_B T = 65$~eV, the boundary between the bands is at $\sim 70$~eV. A genuine lensing event should brighten both bands together, while a flare or instrumental artifact would usually affect just one. Because the magnification itself depends on energy near $w \sim 1$, the test is less reliable there and works best for heavier PBHs.

\section{Limits on PBHs from UV Microlensing}
\label{sec:rates}
In this section we turn the formalism of Sec.~\ref{sec:theory} into expected event rates, and use these rates to set upper limits on $f_{\rm PBH}$. The calculation has three steps. First, we write the event rate per source in terms of the PBH mass and the dark matter fraction. Throughout, we assume that all PBHs have a single mass (see Ref.~\cite{Picker:2026nuu} for extended mass distributions) and that their velocities follow a Maxwell--Boltzmann distribution (see Ref.~\cite{Bozorgnia:2025lsl} for how the Large Magellanic Cloud alters its high-speed tail). In the second step, we describe how the dark matter density changes along each line of sight, and show how the short distances to EUV sources lower the event rate. In the third step, we add up the rate over every event duration we are able to observe, which gives the total number of events we expect to see. Because the EUVE data contain no events, this expected number leads directly to 90\% C.L.\ upper limits on $f_{\rm PBH}$. In Sec.~\ref{sec:montecarlo}, we compare this semi-analytic rate against a Monte Carlo simulation, which confirms that our limits are conservative.

The microlensing rate per source, per unit event timescale $t_E$, is~\cite{Griest1991,Tamta2024}
\begin{align}
\frac{d\Gamma}{dt_E} &= f_{\rm PBH}\,\frac{2D_S}{M v_0^2}\int_0^1 dx\,
\rho_{\rm DM}(x)\,v_E^4(x)\,e^{-v_E^2/v_0^2} \nonumber \\
&\quad \times \int_0^{y_T(x)} \frac{dy}{\sqrt{y_T^2 - y^2}}\;\cdot\;
P(n_\sigma) \,,
\label{eq:dGamma}
\end{align}
where $v_E \equiv 2r_E\sqrt{y_T^2 - y^2}/t_E$ is the transverse lens velocity, $v_0 \simeq 240$~km/s is the halo circular velocity, and $y_T(x)$ is the largest impact parameter for which $\bar{\mu}_E \geq \bar{\mu}_{\rm thresh}$. The factor $P(n_\sigma) = [1 - \Phi(n_\sigma)]^{N_{\rm consec}}$ is the fraction of observing time in which the source count rate stays above the baseline level for $N_{\rm consec}$ consecutive bins. It accounts for the fact that an event can only be detected while the quiescent flux is sufficiently stable. When the counts per bin are low enough that Poisson statistics apply, we replace $P$ by its Poisson counterpart from Eq.~(\ref{eq:Ppoisson}). For the dark matter density profile we use the Einasto form~\cite{Ou2024},
\begin{equation}
\rho_{\rm Ein}(r) = \frac{M_0}{4\pi r_s^3}
\exp\!\left[-\left(\frac{r}{r_s}\right)^\alpha\right] \,,
\label{eq:einasto}
\end{equation}
with parameters $M_0 = 6.2 \times 10^{10}\,M_\odot$, $r_s = 3.86$~kpc, and $\alpha = 0.91$. These values come from a fit to combined Gaia~DR3 and SDSS-APOGEE kinematic data~\cite{Ou2024}. A lens located at fractional position $x$ along the line of sight lies at a galactocentric distance
\begin{equation}
r(x) = \sqrt{r_\odot^2 + x^2 D_S^2 - 2 r_\odot x D_S \cos\ell\cos b} \,,
\end{equation}
where $r_\odot = 8.33$~kpc is the Sun's distance from the Galactic center and $(\ell, b)$ are the Galactic coordinates of the source.

The primary limitation of EUV microlensing relative to X-ray searches is the 
proximity of available sources. For PSR~B0656+14 at $D_S = 290$~pc, the 
line-of-sight optical depth is approximately $10^5$ times smaller than for 
SMC~X-1 at 64~kpc, reflecting the $\tau \propto D_S^2$ scaling at fixed PBH 
mass. We have found that this geometric suppression of the event rate is the primary factor 
restricting the sensitivity of an EUV microlensing search, relative to optical microlensing searches.

The total expected number of microlensing events is
\begin{equation}
N_{\rm ev} = \sum_i T_{\rm obs}^i \int_{t_{\rm min}}^{t_{\rm max}} dt_E\,
\frac{d\Gamma_i}{dt_E} \,,
\label{eq:Nev}
\end{equation}
where the sum runs over sources and $t_{\rm min} = t_{\rm bin}$ along with 
$t_{\rm max} = T_{\rm obs}$ bracket the range of observable timescales. 
In the absence of detected events, a 90\% C.L.\ upper limit on $f_{\rm PBH}$ 
follows from the Poisson condition $N_{\rm ev} = 2.3$.

All integrals are evaluated numerically. We evaluate the $t_E$ integral in Eq.~(\ref{eq:Nev}) using 40-node Gauss--Legendre quadrature in $\log_{10} t_E$. For the $x$ integral in Eq.~(\ref{eq:dGamma}) we apply the trapezoidal rule on a uniform grid of 34 points across $(0,1)$, while the inner $y$ integral is done by adaptive quadrature. The energy average in Eq.~(\ref{eq:muE}) is likewise computed with the trapezoidal rule. To evaluate the confluent hypergeometric function ${}_1F_1$, we use SciPy's \texttt{hyp1f1} wherever it is numerically stable. In the intermediate wave-optics regime, $w = \mathcal{O}(1$--$10)$, where the series for ${}_1F_1$ converges slowly, we switch instead to arbitrary-precision arithmetic with \texttt{mpmath}~\cite{mpmath}.

\subsection{Monte Carlo injection-recovery}
\label{sec:montecarlo}

The event rate above is semi-analytic. The lensing geometry enters through the threshold impact parameter $y_T$, defined by $\bar{\mu}_E(y_T) = \bar{\mu}_{\rm thresh}$, and the chance of detecting an event enters through a single acceptance factor $P(n_\sigma)$ in Eq.~(\ref{eq:dGamma}). As an independent check that avoids the approximations in this factor, we also estimate $N_{\rm ev}$ by injecting simulated lensing events into mock light curves and counting how many are recovered. For a lens mass $M$ and abundance $f_{\rm PBH}$, the expected number of lenses that pass within a generous impact parameter $y_{\rm max}$ of the source (chosen so that $\bar{\mu}_E(y_{\rm max}) < \bar{\mu}_{\rm thresh}$ on every sightline) is
\begin{align}
\lambda &= f_{\rm PBH}\, T_{\rm obs}\, \frac{\sqrt{\pi}\, v_0\, D_S}{M}\, y_{\rm max} \nonumber \\
&\quad \times \int_0^1 \! dx\, \rho_{\rm DM}(x)\, r_E(x) \,,
\label{eq:mc_lambda}
\end{align}
which is simply Eq.~(\ref{eq:dGamma}) integrated over the velocity distribution, with $y_T$ replaced by $y_{\rm max}$ and the acceptance factor left out. From this population we draw a large sample of simulated transits. For each simulated transit, we draw the lens position along the line of sight from $x \sim \rho_{\rm DM}(x)\, r_E(x)$, the distance of closest approach uniformly from $y_0 \sim U[0, y_{\rm max}]$, and the transverse speed from the transit-weighted Maxwellian $P(v) \propto v^2 \exp(-v^2/v_0^2)$. These three quantities together fix the Einstein-crossing time, $t_E = r_E/v$. We then move the lens along its straight-line path, $y(t) = [y_0^2 + (t/t_E)^2]^{1/2}$, and compute the energy-averaged magnification $\bar{\mu}_E(y(t))$ at each point along it. This magnification is sampled in bins of width $t_{\rm bin}$ and added to a simulated light curve whose bins scatter randomly about the magnified flux with fractional amplitude $\sigma_B/B$. We count a simulated event as recovered if it passes the same detection test used in the analytic treatment, that is, $N_{\rm consec}$ consecutive bins above $B + N_\sigma\, \sigma_B$. To obtain the expected number of detected events, we multiply the number of transits, $\lambda$, by the fraction of them that pass the detection test, $\langle f_{\rm rec}\rangle$:
$N_{\rm ev} = \lambda\, \langle f_{\rm rec}\rangle$. As in the analytic case, the 90\% C.L.\ limit corresponds to $N_{\rm ev} = 2.3$.

We test the Monte Carlo in two ways. In the first test, we make it follow the analytic rules: every event with $y_0 < y_T$ is counted, and each is weighted by the flat acceptance $P(n_\sigma)$. The two methods then agree to within the sampling error, which confirms that the overall normalization of the event rate is correct. In the second test, we run the full injection-recovery procedure, in which each event is judged by its own simulated light curve. This gives limits $\sim 1.5$--$2$ times stronger than the analytic ones.

The difference arises because $P(n_\sigma)$ treats every event as if it only barely clears the threshold. In reality, most transits pass close to the source, where $\bar{\mu}_E \gg \bar{\mu}_{\rm thresh}$, and these events are recovered almost every time. The analytic method therefore underestimates the number of detected events, which makes it conservative. We use it for our main limits, and treat the Monte Carlo as an independent check suggesting that the true limits are somewhat stronger.

\section{UV PBH-lensing Results}
\label{sec:results}
This section presents our 90\% C.L.\ upper limits on $f_{\rm PBH}$ from archival EUVE data, along with a projection for CASTOR and a comparison with existing X-ray microlensing limits. Figure~\ref{fig:combined} shows all of these together, with solid lines for limits from existing data and dashed lines for projections.
\subsection{New EUV constraints from PSR~B0656+14}
\label{sec:euvconstraints}
\begin{figure*}[t]
\centering
\includegraphics[width=\textwidth]{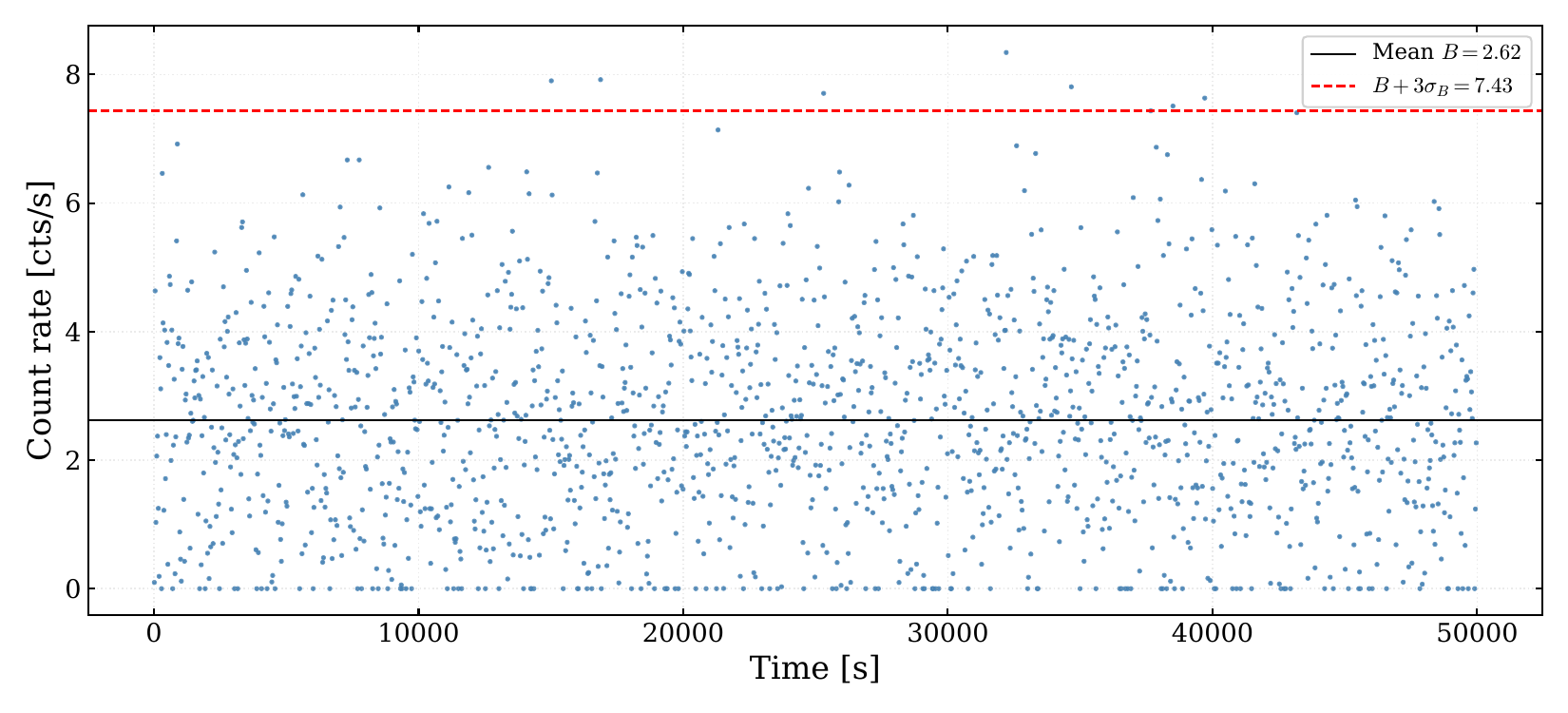}
\caption{EUVE Deep Survey light curve of PSR~B0656+14 in 30~s bins. The solid horizontal line indicates the mean count rate, $B \approx 2.5$~cts/s, and the dashed line indicates the detection threshold, $B + N_\sigma\sigma_B$ with $N_\sigma = 3$. The fractional variability, $\sigma_B/B \approx 0.5$, corresponds to a threshold magnification of $\bar{\mu}_{\rm thresh} = 5.0$. Since no candidate events meet the three-consecutive-bin criterion, the requirement is relaxed to two consecutive bins for this source. The baseline rate is relatively steady, without prominent flares, making PSR~B0656+14 a good target for a microlensing search.}
\label{fig:lightcurve}
\end{figure*}

Figure~\ref{fig:lightcurve} shows the EUVE Deep Survey light curve of PSR~B0656+14, binned in 30~s intervals. The pulsar has a mean count rate of $B \approx 2.5$~cts/s, and its light curve is quite noisy, with a fractional variability of $\sigma_B/B \approx 0.50$. No candidate events pass our standard requirement of three consecutive bins above the $3\sigma$ threshold. Because the light curve is so noisy, we relax this requirement to two consecutive bins for this source.

The resulting 90\%~C.L.\ upper limit on $f_{\rm PBH}$, calculated with the Einasto dark matter profile, is shown as the \textcolor{blue}{blue} shaded region in Fig.~\ref{fig:combined}. It spans the mass range $\sim 5\times10^{-13}$--$3\times10^{-12}\,M_\odot$. Its lower edge comes from wave optics: below this mass, $w \lesssim 1$ even at the highest EUVE energy, $E_{\rm max} = 124$~eV, and diffraction removes the lensing signal. The strongest limit, $f_{\rm PBH} \approx 120$ near $M \approx 1.2\times10^{-12}\,M_\odot$, is well above one, so it does not yet constrain the PBH abundance. The main reason the limit is weak is that PSR~B0656+14 is very close to us. Because the pulsar lies only $\sim290$~pc away, there is little dark matter along the line of sight, and the lensing optical depth is about $10^5$ times smaller than toward SMC~X-1. At high masses, the limit weakens as $f_{\rm PBH} \propto \sqrt{M}$. The reason is that, for a fixed dark matter density, heavier PBHs are fewer in number, and this drop in number outweighs the growth of their Einstein radius.

\subsection{New EUV constraints from the white dwarf HZ~43}
\label{sec:finitesource}
\begin{figure*}[t]
\centering
\includegraphics[width=\textwidth]{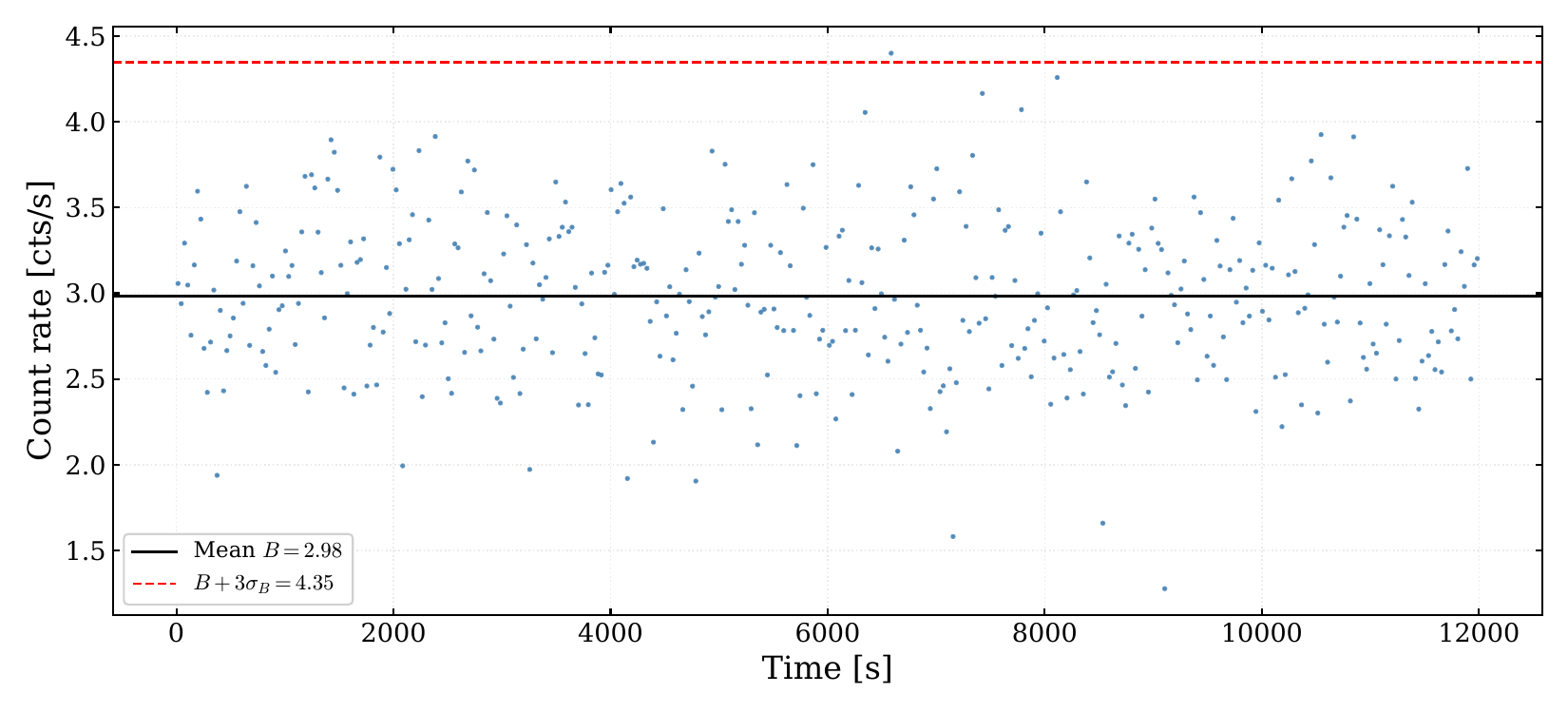}
\caption{EUVE Deep Survey light curve of the white dwarf HZ~43 (50~h net exposure), binned at 30~s. Lines are as in Fig.~\ref{fig:lightcurve}. HZ~43 is bright ($B \approx 3.0$~cts/s), and at this binning its fractional variability, $\sigma_B/B \approx 0.11$, is dominated by photon counting noise, giving a threshold magnification $\bar{\mu}_{\rm thresh} \approx 1.5$. No candidate microlensing events are identified.}
\label{fig:lightcurve_hz43}
\end{figure*}

\begin{figure*}[t]
\centering
\includegraphics[width=0.9\textwidth]{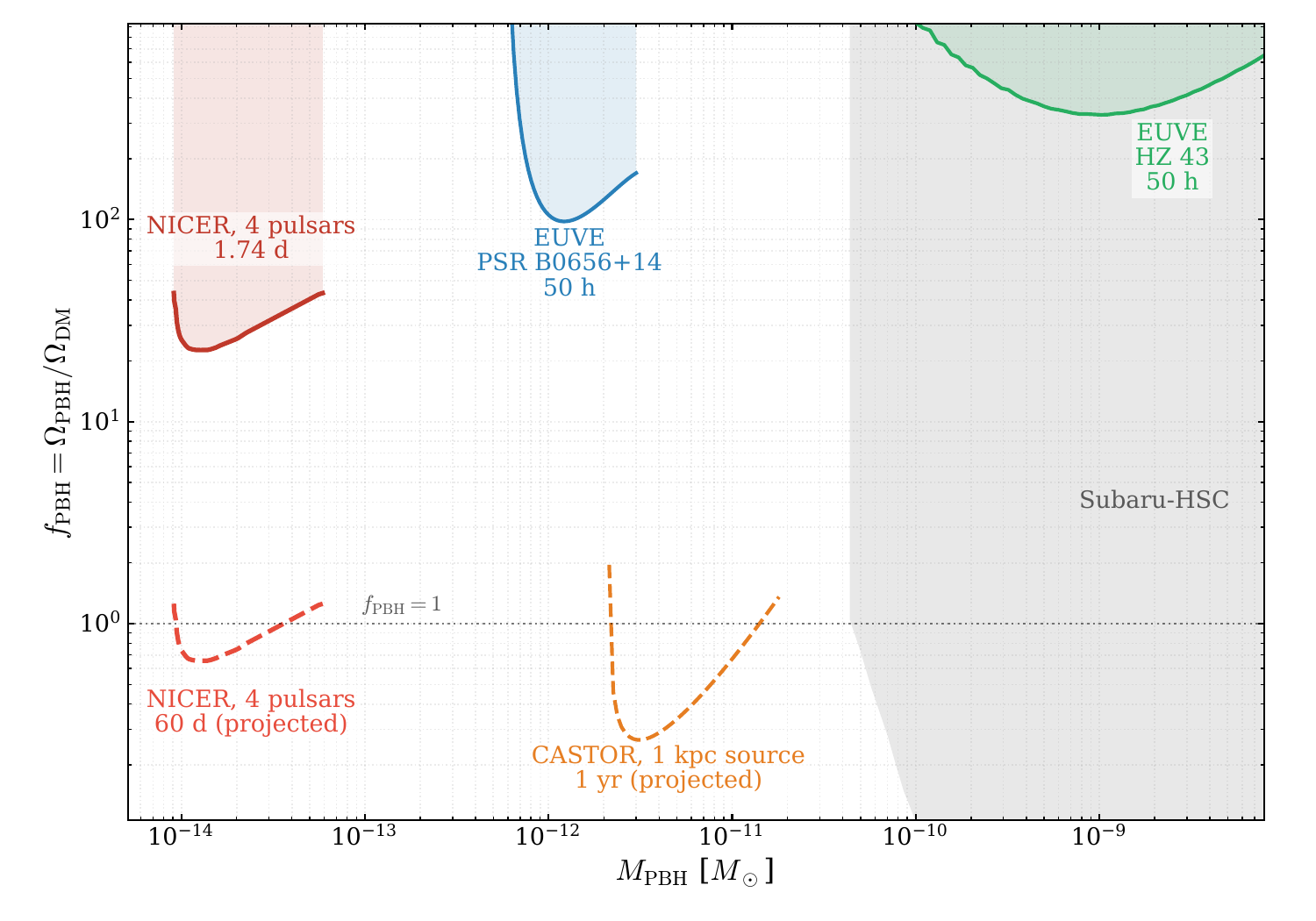}
\caption{90\%~C.L.\ upper limits on the dark matter fraction of asteroid-mass PBHs. Solid lines represent constraints from existing data, whereas, the dashed line are future projections. \textcolor{blue}{Blue}: EUVE, PSR~B0656+14, 50h (Sec.~\ref{sec:euvconstraints}). \textcolor{forestgreen}{Green}: EUVE, HZ~43, 50h (Sec.~\ref{sec:finitesource}). \textcolor{orange}{Orange}: CASTOR, compact source at 1~kpc, 1yr (Sec.~\ref{sec:castor}). \textcolor{red}{Red}: NICER, four X-ray pulsars of Table~I of Ref.~\cite{Tamta2024}, 1.74~days (solid) and a 60-day projection (dashed) (Sec.~\ref{sec:xrayvalidation}). For comparison, we overlay (\textcolor{gray}{gray}) the Subaru-HSC optical microlensing constraint of M31~\cite{Niikura2019,Smyth:2019whb}, which excludes more massive PBHs as a dominant component of dark matter.}
\label{fig:combined}
\end{figure*}

Figure~\ref{fig:combined} shows the limit from HZ~43, one of the brightest white dwarfs in the EUVE catalog, as the \textcolor{ForestGreen}{green} shaded region. Its light curve, based on 50~hours of EUVE Deep Survey exposure, is shown in Fig.~\ref{fig:lightcurve_hz43}. HZ~43 is a clear example of how a large source size limits sensitivity. Even at $M = 10^{-10}\,M_\odot$, two orders of magnitude above the mass where the pulsar limit is strongest, it is too large to be noticeably magnified: for a lens passing at impact parameter $y = 0.1$, its energy-averaged magnification is just $\bar{\mu}_E \approx 1.02$, far below any practical detection threshold. O/B stars do even worse. Their projected sizes satisfy $a_S \gtrsim 10^4$, which keeps their magnification essentially at one over the whole mass range. They therefore provide no constraint, and we leave them out of Fig.~\ref{fig:combined}.

The way the HZ~43 limit changes with mass is still instructive. Since the Einstein radius grows with mass, $a_S$ falls as $M^{-1/2}$. The suppression only disappears once $r_E$ becomes comparable to the projected size of the source, which requires PBH masses 10 to 100 times larger than those where the pulsar limit is strongest.

At these higher masses, HZ~43 benefits from its low variability. With 30~s bins, its fractional variability is $\sigma_B/B \approx 0.11$, mostly from photon counting, which is about five times lower than that of PSR~B0656+14 ($\sigma_B/B \approx 0.50$). This lowers the detection threshold from $\bar{\mu}_{\rm thresh} \approx 5.0$ to $\approx 1.5$, so a lens can produce a detectable event even when it passes farther from the source. The \textcolor{ForestGreen}{green} curve therefore begins at $M \gtrsim 10^{-10}\,M_\odot$ and improves steadily with mass, reaching $f_{\rm PBH} \approx 1.4\times10^{3}$ at $M \approx 8\times10^{-9}\,M_\odot$, the top of our mass range. The lower threshold still cannot make up for the small optical depth toward such a nearby source, and the limit stays an order of magnitude above the pulsar's $f_{\rm PBH} \approx 120$.

In addition, the whole HZ~43 curve lies inside the region already excluded by Subaru-HSC optical microlensing~\cite{Niikura2019}, which reaches $f_{\rm PBH} \sim 0.03$ near $10^{-9}\,M_\odot$, so it adds no new exclusion. PSR~B0656+14, by contrast, is the only EUV source whose limit lies in the open mass window below the Subaru-HSC range, at $M \sim 5\times10^{-13}$--$3\times10^{-12}\,M_\odot$, which is why it is the EUV result we quote in the rest of this work. The HZ~43 curve is still useful, because it shows finite-source suppression in action: the suppression scales with $a_S$ as the analytic treatment predicts, and lifts only at PBH masses well above the open window of interest. This reinforces a theme that runs through this work, namely that a compact source is a necessary condition for competitive short-wavelength microlensing in the asteroid-mass window. The same requirement also supports the decision of Ref.~\cite{Tamta2024} to restrict their X-ray analysis to neutron-star pulsars.

\subsection{Projected sensitivity with CASTOR telescope}
\label{sec:castor}
Fig.~\ref{fig:combined} (\textcolor{Orange}{orange} dashed line) shows a future sensitivity projection for PBH microlensing with the proposed CASTOR space telescope~\cite{Cote2019,CASTOR_SPIE2025}. CASTOR is a Canadian-led space telescope planned for launch in the 2030s. It would carry a 1-m mirror and observe in the ultraviolet and blue optical, with three imaging channels covering 150--550~nm, fast CMOS detectors capable of rapid photometry, and a wide 0.25~deg$^2$ field of view. Of these channels, the UV channel (150--300~nm, or $E \approx 4$--$8$~eV) is the one most relevant for microlensing. We use the instrument parameters listed in Table~\ref{tab:castor2}, which are drawn from published mission specifications~\cite{Cote2019,CASTOR_SPIE2025,FORECASTOR}.

\begin{table}[h]
\centering
\begin{tabular}{lc}
\hline\hline
Parameter & Value \\
\hline
Primary mirror diameter & 1.0~m \\
UV channel wavelength range & 150--300~nm \\
UV channel energy range & 4.1--8.3~eV \\
Peak effective area (UV) & $\sim 1500$~cm$^2$ \\
Photometry time resolution & $\sim 10$~ms \\
\hline\hline
\end{tabular}
\caption{Assumed CASTOR UV-channel parameters for our projected sensitivity estimate.}
\label{tab:castor2}
\end{table}

The peak effective area of $\sim1500$~cm$^2$ follows from a 1-m unobscured aperture (geometric area 7850~cm$^2$), reduced by the reflectivity of the three-mirror telescope ($\sim73\%$), dichroic and filter transmission ($\sim80\%$), and the quantum efficiency of UV-enhanced, 2D-doped CMOS detectors ($\sim30$--$50\%$)~\cite{Skottfelt2024}. This is roughly $15\times$ the effective area of the EUVE Deep Survey detector.
 
Two factors set CASTOR’s accessible PBH mass range. From below, the wave-optics criterion $w \sim 1$ evaluated at $E_{\rm max} = 8.3$~eV gives a minimum accessible mass $M_{\rm min} \approx 2\times10^{-12}\,M_\odot$, several times higher than for EUVE because CASTOR detects lower-energy photons. From above, the mass range extends to $\sim10^{-11}\,M_\odot$, almost overlapping with the Subaru-HSC optical lensing constraint~\cite{Niikura2019,Smyth:2019whb}.

The key advantage of CASTOR for microlensing lies not only in mass reach but in sensitivity. UV photons at 150--300~nm are far less absorbed by the interstellar medium than EUV photons. The cause of this is the ISM opacity at 200~nm being dominated by dust rather than photoionization, with typical extinctions of $A_{200} \sim 2$--$5$~mag/kpc along low-$N_H$ sightlines. This enables observations of compact UV sources at distances of $\sim 1$~kpc, roughly $3$--$4\times$ farther than the EUVE horizon, with corresponding improvements in microlensing optical depth.

To quantify this improvement, we construct a sensitivity projection by considering CASTOR observations of a compact UV source at $D_S = 1$~kpc, modeled as a thermally-emitting millisecond pulsar with non-thermal UV emission at the level detected from analogous objects by HST~\cite{Kargaltsev2012}. For this source we take a neutron star radius of $R_S = 13$~km and a fractional variability of $\sigma_B/B = 0.15$. From the source's UV luminosity and CASTOR's effective area, we estimate a count rate of 0.5~cts/s, and we assume one year of dedicated monitoring. With these parameters, $\bar{\mu}_{\rm thresh} \approx 1.7$, and the Poisson-corrected detection probability is $P(n_\sigma = -1)^{N_{\rm consec}} = 0.60$ for $\sim5$~counts per 10~s bin. The \textcolor{Orange}{orange} dashed line in Fig.~\ref{fig:combined} shows the resulting projection.

Two improvements drive the gain over EUVE. First, a source at $D_S = 1$~kpc has a $\sim12$ times larger microlensing optical depth, since $\tau \propto D_S^2$ along short sightlines. Second, CASTOR's larger effective area and steadier photometry lower $\bar{\mu}_{\rm thresh}$ by a factor of $\sim4$, so each lens can be detected over a wider area around the line of sight. Together, these give $f_{\rm PBH} \approx 0.35$ near $M \approx 3.5\times10^{-12}\,M_\odot$. This is more than two orders of magnitude below the EUVE limit and, crucially, below one, which is what it takes to exclude PBHs as the main component of dark matter. CASTOR would be the first UV telescope to reach this point, closing part of the open window at $\sim2\times10^{-12}$--$10^{-11}\,M_\odot$ that no existing survey covers.

This projection rests on one key assumption: that bright, compact UV sources exist at $\gtrsim1$~kpc along sightlines with low extinction. No such source has yet been confirmed, so finding one is the main hurdle. Better distances to nearby neutron stars, for instance from new maps of the local electron density~\cite{Bramante:2024ikc}, would help in this search.

\subsection{Comparison with the NICER X-ray limits}
\label{sec:xrayvalidation}
As a check on our pipeline, and to put the EUV results in context, we compare its output with the X-ray microlensing analysis of Ref.~\cite{Tamta2024}. Their Fig.~1 gives the NICER limit from 1.74~days of data on four pulsars. Using the same inputs, the Einasto profile, and the NICER-XTI effective area~\cite{Wilson-Hodge2016}, we reproduce their limit to within a few percent at all masses, including the low-mass wave-optics cutoff and the high-mass $f_{\rm PBH} \propto \sqrt{M}$ rise. These are the \textcolor{red}{red} curves in Fig.~\ref{fig:combined}. The solid line, from the existing data, reaches $f_{\rm PBH} \approx 23$ near $M \approx 1.4\times10^{-14}\,M_\odot$. The dashed line scales this to 60 days, since $N_{\rm ev} \propto T_{\rm obs}$, and reaches $f_{\rm PBH} \approx 0.66$, enough to rule out PBHs as the dominant dark matter at those masses.

\subsection{Reach for finite-size dark compact objects}
\label{sec:massradius}
So far we have treated the lens as a point mass, which is an excellent approximation for a PBH. The same data, however, also constrain dark compact objects (DCOs) whose mass $M_L$ and radius $R_L$ are independent~\cite{Croon:2020wpr,Fairbairn2017,BaiLongLu2020}. Examples include DCOs formed by dissipative dynamics in a dark sector~\cite{Bramante:2023ddr,Bramante:2024pyc}, composite dark matter states~\cite{Bramante:2026wzh,Acevedo:2024lyr,Bleau:2025klr}, and boson stars~\cite{Acevedo:2022gug}. Such objects could also be detected through the dimming of background starlight~\cite{Bramante:2024hbr}.

Whether microlensing can detect a DCO depends on how compact it is. A DCO larger than its own Einstein radius is too diffuse to focus light well, so its magnification drops once $R_L \gtrsim r_E(M_L)$. Wave optics, by contrast, limits the lens mass rather than its size. Geometric optics applies once $w = 4GM_L\omega/c^3 \gtrsim 1$, a condition that depends on the photon frequency and the lens mass but not on $R_L$. Above this mass, any lens lying between its Schwarzschild radius and its Einstein radius lenses efficiently.

\begin{figure*}[t]
\centering
\includegraphics[width=0.9\textwidth]{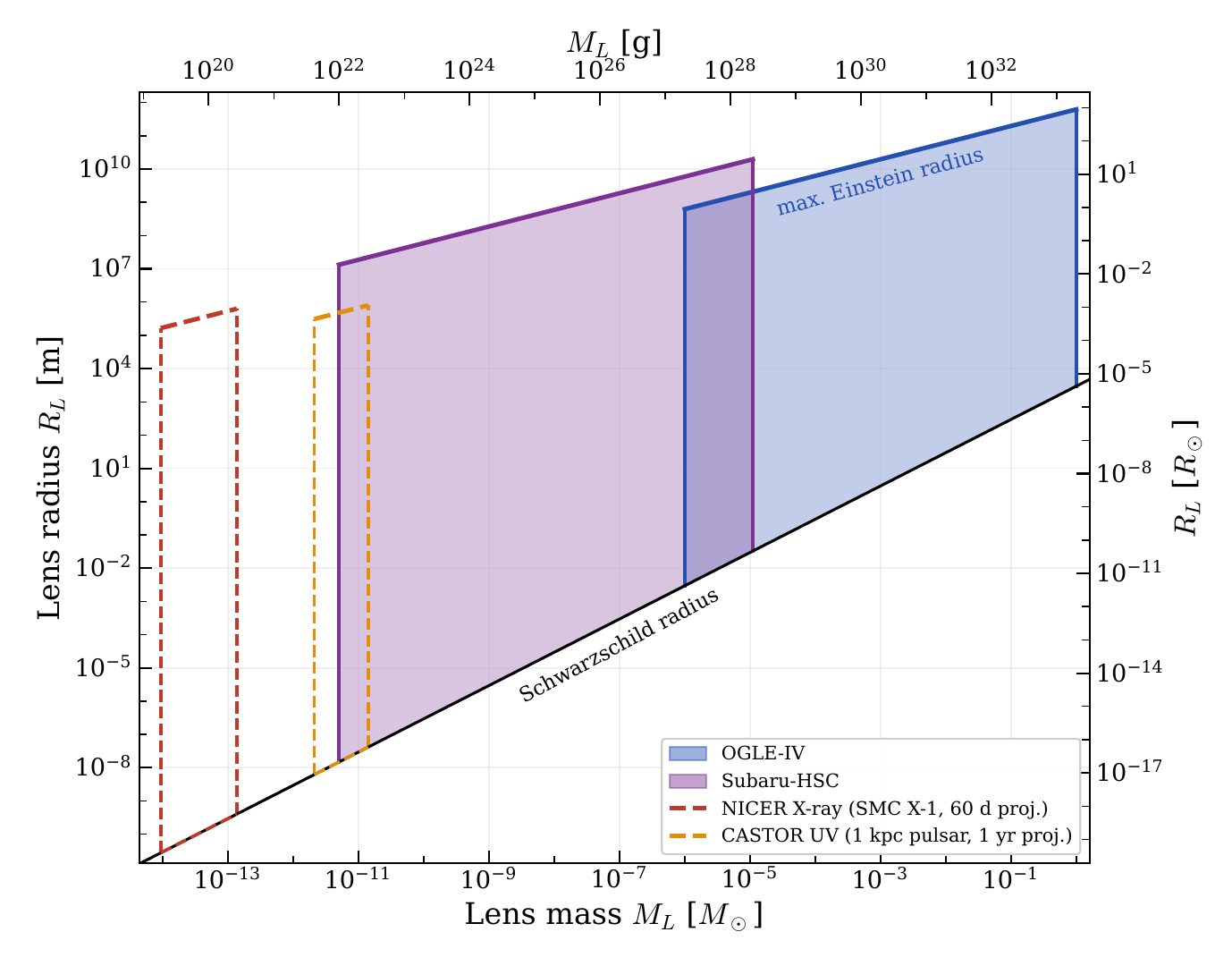}
\caption{Reach of microlensing probes in the plane of lens mass $M_L$ and lens radius $R_L$. Each region marks the range of objects that a survey is able to probe. The shaded regions are from existing data, and the dashed outlines are projections. Each region is bounded above by the Einstein radius $r_E(M_L)$, since larger lenses are too diffuse to magnify, and below by the Schwarzschild radius $R_S = 2GM_L/c^2$ (black diagonal line), since no lens can be denser than a black hole. 
The low-mass edge of each region is set by the wave-optics cutoff, which shifts to lower mass for higher photon energy. Shown are
the optical OGLE-IV survey of the Galactic bulge (\textcolor{blue}{blue}) and Subaru-HSC survey of M31 (\textcolor{purple}{purple}), following Ref.~\cite{Croon2020}, and our projected NICER X-ray (\textcolor{red}{red}, 60-day SMC~X-1) and CASTOR UV (\textcolor{orange}{orange}, one-year, 1~kpc source) probes.  Right and top axes give $R_L$
in solar radii and $M_L$ in grams.}
\label{fig:massradius}
\end{figure*}

Each survey is therefore sensitive to a region of the $(M_L, R_L)$ plane. The region is bounded above by the Einstein radius, $r_E(M_L) \propto \sqrt{M_L}$, and below by the Schwarzschild radius, $R_S = 2GM_L/c^2$, since nothing can be more compact than a black hole. On the low-mass side it is bounded by the wave-optics cutoff, which moves to lower masses at higher photon energies. Figure~\ref{fig:massradius} shows these regions, in the style of Fig.~1 of Ref.~\cite{Croon:2020wpr}, for two optical surveys, OGLE-IV toward the Galactic bulge~\cite{Croon:2020wpr} and Subaru-HSC toward M31~\cite{Croon2020}, and for our projected NICER X-ray and CASTOR UV searches.

The probes cover different ranges of mass and size. The optical surveys reach the largest lenses. OGLE-IV extends up to $r_E \sim 10^{12}$~m ($\sim10^3\,R_\odot$) near a solar mass, but only for $M_L \gtrsim 10^{-6}\,M_\odot$. Subaru-HSC, whose M31 sources lie much farther away, covers $M_L \sim 4\times10^{-11}$--$8\times10^{-6}\,M_\odot$, with $r_E$ up to $\sim 10^{10}$~m ($\sim 20\,R_\odot$). At OGLE-IV masses the Schwarzschild radius is much larger than the optical wavelength, so wave effects play no role and the lower size bound is simply $R_S$. The X-ray and UV probes instead reach the low masses where PBHs could still make up all of the dark matter. NICER covers $M_L \sim 10^{-14}$--$10^{-13}\,M_\odot$ and CASTOR covers $M_L \sim 10^{-12}$--$10^{-11}\,M_\odot$, both up to $r_E \sim 10^5$--$10^6$~m. Because higher-energy photons push the wave-optics cutoff to lower masses, NICER reaches the smallest masses, $M_L \sim 10^{-14}\,M_\odot$, where the Schwarzschild radius is only $R_S \sim 3\times10^{-11}$~m. It can therefore test DCOs that are nearly as compact as black holes, at masses that optical and UV surveys cannot reach. The upper boundary shown in the figure is the Einstein radius of a point-like lens~\cite{Croon:2020wpr}.

\section{Conclusions}
\label{sec:discussion}
This work has shown that UV microlensing is a potentially powerful probe of PBH dark matter in the asteroid-mass 
window, complementary to existing optical and proposed X-ray searches. In this work we have applied a finite-source, wave-optics microlensing formalism to EUV sources for the first time, and obtained the first EUV microlensing bound on PBHs. Our main finding is that EUV observations of nearby neutron stars can probe dark matter in the mass range $\sim 10^{-13}$--$10^{-11}\,\Msun$, even though interstellar absorption limits how distant such sources can be. With PSR~B0656+14 ($D_S = 290$~pc), the best EUV target currently available, we obtain 90\%~C.L.\ upper limits that reach $f_{\rm PBH} \approx 120$ at $M_{\rm PBH} \approx 1.2 \times 10^{-12}\,\Msun$. Although this is not yet strong enough to rule out PBHs as the dark matter, it establishes the method and the target-selection criteria for future searches.

Our second result is that EUV microlensing depends on compact sources. At the lowest masses we probe, only thermally emitting neutron stars, whose emission regions have $R_S \sim 10$--$15$~km, are small enough to escape severe finite-source suppression. White dwarfs, with $R_S \sim 7000$~km, remain suppressed across most of our mass range. Once the suppression eases, at $M \gtrsim 10^{-10}\,\Msun$, HZ~43 reaches at best $f_{\rm PBH} \approx 1.4\times10^{3}$, within the region already excluded by Subaru-HSC. O/B stars are suppressed at every mass we consider. Because this ordering of source types applies to short-wavelength microlensing generally, future UV campaigns should concentrate on compact objects.

Looking ahead, a next-generation UV telescope such as CASTOR could reach $f_{\rm PBH} \approx 0.35$ near $M \approx 3.5 \times 10^{-12}\,\Msun$ with one year of observing, and could stay below one over roughly a decade in mass ($\sim 2 \times 10^{-12}$--$10^{-11}\,\Msun$). This would be the first short-wavelength microlensing exclusion inside the open asteroid-mass window. The improvement over EUVE, more than two orders of magnitude, comes from three factors. First, UV light at 150--300~nm is far less absorbed by the interstellar medium, which opens up sightlines to compact sources at $\sim\!1$~kpc. Second, CASTOR's $\sim\!15\times$ larger effective area collects more photons. Third, the resulting better photon statistics lower the detection threshold. Taken together, this projection makes a strong case for including the monitoring of compact sources as an explicit PBH science goal of upcoming UV missions such as ULTRASAT and UVEX~\cite{Shvartzvald2024,Kulkarni:2021tit}.

PBH and DCO microlensing at any wavelength benefits from three key source properties: compactness ($R_S \lesssim r_E$), distance ($D_S$ as large as possible for optical depth), and brightness (low $\sigma_B/B$ for a low detection threshold $\bar{\mu}_{\rm thresh}$). Sources satisfying all three criteria at UV/EUV wavelengths are rare, yet the rapid growth of time-domain UV astronomy — driven by missions such as CASTOR, ULTRASAT, and UVEX — may reveal new targets enabling competitive microlensing searches at wavelengths shorter than X-ray. We therefore encourage the community to consider dedicated monitoring of compact UV sources as an explicit PBH dark matter science case in the planning of these missions.

\section*{acknowledgments}
We sincerely thank Zac Picker for helpful conversations, and Nirmal Raj and Manish Tamta for very useful correspondence on PBH X-ray lensing. We acknowledge support from the Natural Sciences and Engineering Research Council of Canada (NSERC), the Ontario Early Researcher Award (ERA), and the Canada Foundation for Innovation (CFI). This research was undertaken thanks in part to funding from the Arthur B. McDonald Canadian Astroparticle Physics Research Institute. Research at Perimeter Institute is supported
by the Government of Canada through the Department of Innovation, Science, and Economic Development, and by the Province of Ontario.

\bibliographystyle{apsrev4-2}
\bibliography{asteroid}

\end{document}